\documentclass[english,twocolumn,prx,superscriptaddress]{revtex4}
\usepackage[T1]{fontenc}
\usepackage[latin9]{inputenc}
\usepackage{amstext}
\usepackage{graphicx}
\usepackage{esint}

\makeatletter
\@ifundefined{textcolor}{}
{%
 \definecolor{BLACK}{gray}{0}
 \definecolor{WHITE}{gray}{1}
 \definecolor{RED}{rgb}{1,0,0}
 \definecolor{GREEN}{rgb}{0,1,0}
 \definecolor{BLUE}{rgb}{0,0,1}
 \definecolor{CYAN}{cmyk}{1,0,0,0}
 \definecolor{MAGENTA}{cmyk}{0,1,0,0}
 \definecolor{YELLOW}{cmyk}{0,0,1,0}
}

\makeatother

\usepackage{babel}
\begin{document}

\title{Effects of system-bath entanglement  on the performance
of light-harvesting  systems: A quantum heat engine perspective}

\author{Dazhi Xu }

\affiliation{School of Materials Science and Engineering, Nanyang Technological
University, Singapore }

\affiliation{Department of Chemistry, Massachusetts Institute of Technology, Cambridge,
MA 02139 U.S.A }

\author{Chen Wang}

\affiliation{Department of Chemistry, Massachusetts Institute of Technology, Cambridge,
MA 02139 U.S.A }

\affiliation{Singapore-MIT Alliance for Research and Technology, 1 CREATE Way,
Singapore 138602, Singapore }

\author{Yang Zhao}

\affiliation{School of Materials Science and Engineering, Nanyang Technological
University, Singapore }

\author{Jianshu Cao}

\email{jianshu@mit.edu}
\affiliation{Department of Chemistry, Massachusetts Institute of Technology, Cambridge,
MA 02139 U.S.A }

\affiliation{Singapore-MIT Alliance for Research and Technology, 1 CREATE Way,
Singapore 138602, Singapore }

\begin{abstract}
We explore energy transfer in a generic three-level system, which is coupled
to three non-equilibrium baths.  Built on the concept of quantum heat engine, 
 our three-level model describes non-equilibrium quantum processes including
  light-harvesting energy transfer, nano-scale heat transfer, photo-induced isomerization,
 and photovoltaics in double quantum-dots. In the context of light-harvesting,
 the excitation energy is first pumped up by sunlight,  then is transferred via two excited
states which are coupled to a phonon bath, and finally decays to the
ground state. The efficiency of this process is evaluated
by steady state analysis via a polaron-transformed master equation;
thus a wide range of the system-phonon coupling strength can
be covered. We show that the coupling with the phonon bath not only
modifies the steady state, resulting in population inversion, but
also introduces a finite steady state coherence which optimizes 
the energy transfer flux and efficiency. In the strong coupling limit, the
steady state coherence disappears and the efficiency approaches 
the heat engine limit given by Scovil and Schultz-Dubois  
in  Phys. Rew. Lett. \textbf{2},  262 (1959).
\end{abstract}
\maketitle

\section{Introduction}

With the rapid developments in measurement and manipulation of microscopic
systems, quantum effects such as coherence and entanglement are often
utilized to enhance the performance of microscopic devices. Even in
biological systems, both experiments \cite{Fleming2007} and theoretical
models \cite{Ishizaki2009,JLWu2010} reveal that the long-lived quantum
coherence may play an important role in highly efficient energy and
electron transfer processes. How biological systems, such as light-harvesting
complex, preserve such long-lived coherence and how nature benefits
from the coherence are two key questions that define the emerging field of quantum biology. 

Taking a three-level system as a generic theoretical model, many interesting
mechanisms can be well demonstrated and understood. Recently, the sunlight-induced 
exciton coherence  is studied in
a V-configuration three-level model \cite{Brumer2014,Dijkstra2014}.
An interesting idea is to consider the energy transfer process from
the perspective of heat engine \cite{Quan2007}. 
For example, the coherence introduced by an auxiliary energy level
can enhance the heat engine power \cite{Scully2011,Scully2011-2}. 
The early work considering a three-level
maser model as a Carnot engine was carried out by Scovil and Schulz-DuBois
\cite{SSD1959,SSD1967}, yielding the heat engine efficiency $\eta_{0}$
and its relation with the Carnot efficiency. Later papers
elaborately reexamined the dynamics of this model by the Lindblad master
equation and showed that the thermodynamic efficiency $\eta_{0}$
is achieved when the output light-field is strongly coupled with the
three-level system \cite{Kosloff1996,Tannor2006,Tannor2007}. The quantum heat engine provides us a heuristic
perspective to better understand the basic physical processes in energy
transfer and presents useful insight to enhance the efficiency and
output power in small systems \cite{Popescu2010,Kosloff2012,Mukamel2012,Adesso2014}.

In this paper, we study the polaron effects of a phonon bath on the
energy transfer flux and efficiency in a generic three-level model. The canonical
distribution of a thermal equilibrium system requires a negligible coupling between
the system and its environment. As the coupling strength grows, the
steady state of the system will no longer be canonical \cite{Dong2007,Dong2010,JSCso2012,Cao2012,Xu2014}.
This non-canonical state actually introduces the steady state coherence
into the system without refereeing to specific forms of light-matter
interaction or  designing exotic system configurations. The bath-induced coherent
effect is investigated by the polaron-transformed Redfield equation
(PTRE) \cite{Silbey1971,Silbey1984}, which bridges both the weak and
strong system-bath coupling regions. The difference between the steady
state efficiency and strong coupling limit  $\eta_{0}$
depends strongly on the phonon-induced coherence. Taking into
account of the behaviour of both the flux and efficiency, we are able
to optimize coupling and temperature in designing optimal
artificial energy transfer systems. 

In this paper, we first introduce the three-level model and its non-equilibrium
environment in section II, and then formulate  the PTRE in section
III. In section IV, the polaron effects of phonon-bath on the energy
transfer flux and efficiency are studied in detail. We summarize our
results in the last section.

\section{three-level system model}

\subsection{Model system}

\begin{figure}
\includegraphics[width=6cm]{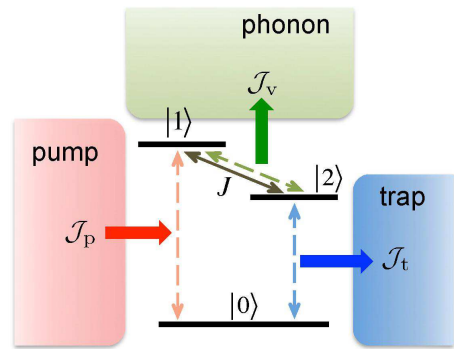}

\protect\caption{\label{fig:scheme}(color online) The system is modeled by a three-level
system: its ground state $\left|0\right\rangle$ and the excited
state $\left|1\right\rangle $ ($\left|2\right\rangle $) is coupled
with the pumping (trapping) bath; the excited states $\left|1\right\rangle$
and $\left|2\right\rangle$ are diagonal-coupled with the phonon
bath; the internal transition strength between $\left|1\right\rangle$
and $\left|2\right\rangle $ is characterized by $J$. The energy
fluxes $\mathcal{J}_{\textrm{p}}$, $\mathcal{J}_{\textrm{v}}$ and
$\mathcal{J}_{\textrm{t}}$ describe the energy exchange rate of the
system with the pumping, the phonon and the trapping baths, respectively.
The flux into the system is defined as the positive direction.}
\end{figure}

We consider the energy transfer process in the three-level system
illustrated in Fig.\ref{fig:scheme}. The site energy of the ground
state $\left|0\right\rangle$ is set to zero. The two excited energy
levels $\left|1\right\rangle$ and $\left|2\right\rangle$ form
a two-level system (TLS, in the following the TLS is referred to the
two excited states), with the corresponding site energy $\epsilon_{1}$
and $\epsilon_{2}$. The transition due to the dipole-dipole interaction
is characterized by $J$. Then the three-level system is modeled by
the Hamiltonian $H_{0}$ as: 
\begin{eqnarray}
H_{0} & = & \sum_{i=1,2}\epsilon_{i}\left|i\right\rangle \left\langle i\right|+\frac{J}{2}\left(\left|1\right\rangle \left\langle 2\right|+\left|2\right\rangle \left\langle 1\right|\right).\label{eq:H0}
\end{eqnarray}
We are interested in the transfer process in the single excitation
subspace: The three-level system is firstly excited to state $\left|1\right\rangle$
by a photon field, then the excitation is transferred to state $\left|2\right\rangle$
through $J$ (mediated by phonon modes),
and finally the excitation decays to the ground state $\left|0\right\rangle$
via spontaneous radiation. The pumping and trapping processes are
modeled by the interaction with the two independent photon baths,
which are coupled separately with two transitions $\left|0\right\rangle \leftrightarrow\left|1\right\rangle$
and $\left|0\right\rangle \leftrightarrow\left|2\right\rangle$.
The Hamiltonian of the photon baths and their interactions with the
three-level system are given by 

\begin{eqnarray}
H_{i=\textrm{p},\textrm{t}} & = & \sum_{k}\omega_{ik}a_{ik}^{\dagger}a_{ik}+\left(g_{ik}a_{ik}^{\dagger}\left|0\right\rangle \left\langle i\right|+\mbox{H.c.}\right),\label{eq:Hpump}
\end{eqnarray}
where $\omega_{ik}$ is the eigen frequency of the bath mode described
by the creation (annihilation) operator $a_{ik}^{\dagger}$ ($a_{ik}$),
and its coupling strength to the excited state is $g_{ik}$. We note
that the rotating wave approximation is applied in the system-bath
interaction term. A phonon bath with creation and annihilation operators
$b_{k}^{\dagger}$ and $b_{k}$ of the bath mode $\omega_{\textrm{v}k}$
is coupled to the TLS via diagonal interaction with the coupling strength
of $f_{k}$. Thus, the phonon part is described by 
\begin{equation}
H_{\textrm{v}}=\sum_{k}\omega_{\textrm{v}k}b_{k}^{\dagger}b_{k}+\left(\left|1\right\rangle \left\langle 1\right|-\left|2\right\rangle \left\langle 2\right|\right)\sum_{k}\left(f_{k}b_{k}^{\dagger}+\mbox{H.c.}\right).\label{eq:Hpho}
\end{equation}

\begin{figure}
\includegraphics[width=8cm]{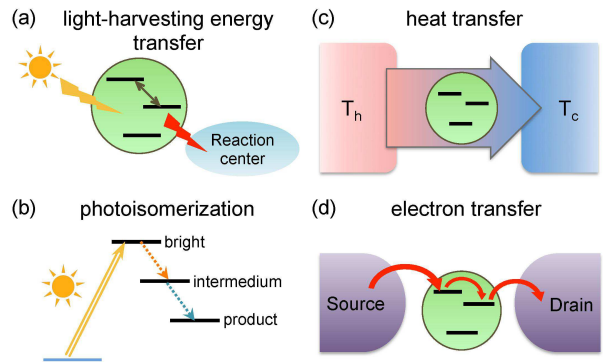}\protect\caption{\label{fig:1-2}
Realistic examples which can be studied by the
three-level model with different heat baths. (a) In the photosynthesis
process, the three-level system works as an antenna that captures
the energy from sunlight and then transfers to the reaction center. (b)
Three eigenstates manifolds in photoisomerization. The bright states
are pumped by the light field, then the populations relax to the intermedium 
and product states in the phonon environment.
(c) The heat transfer in nanoscale can also use the three-level system
as a bridge connecting the high temperature and low temperature heat
baths. (d) In the electron transport problem, electrons tunnel through
double quantum dots which can be described by a three-level system. The
quantum dot connects with a source and a drain.}

\end{figure}

This microscopic three-level system immersed in the non-equilibrium
environment was studied as a quantum heat pump phenomenologically
without considering the details of the system-bath coupling \cite{SSD1967}.
In the case that the phonon bath is replaced by a single driving mode strongly
coupled to the system, the dynamic steady states have been solved
and the efficiency  is given by $\eta_{0}=\epsilon_{2}/\epsilon_{1}$
\cite{Tannor2006,Tannor2007}. In reality, the three-level model can
be realized in both nature and laboratory. Taking
the energy transfer process in photosynthetic pigment for example
[Fig.\ref{fig:1-2}(a)], different baths could arise from different
sources: the pumping light field (such as the sun-light photons) 
is considered as a high temperature boson
bath; the trapping bath is formed by the surrounding electromagnetic
environment which models the energy transfer to the reaction center;
and the phonon bath with inverse temperature $\beta_{\textrm{v}}$
describes the phonon modes coupled with the excited states. In addition,
such a three-level (or more intermediate energy levels) system can
be used to describe photoisomerization [Fig.\ref{fig:1-2}(b)], 
nanoscale heat transfer \cite{ Cao2015} [Fig.\ref{fig:1-2}(c)] 
or photovoltaic current in double quantum dots \cite{CWang2014} [Fig.\ref{fig:1-2}(d)]. 

In this paper, we focus on the effects of the phonon modes on energy flux and efficiency. 
Usually when the system-phonon bath coupling strength is
not weak, the Bloch-Redfield equation approach cannot be applied.
Therefore, we will introduce the polaron-transformed Redfield equation
(PTRE) \cite{Silbey1971,Silbey1984}, which gives reliable results
from the weak to strong coupling region, to study the bath-induced
coherent effects of this quantum system.

\subsection{Definitions of energy flux and transfer efficiency }

We are interested in the energy transfer flux and efficiency of the
three-level system at its non-equilibrium steady state. The steady
state solution can be obtained by the master equation formally written
as 
\begin{equation}
\frac{d\rho_{s}(t)}{dt}=\left(\mathcal{L}_{0}+\mathcal{L}_{\textrm{p}}+\mathcal{L}_{\textrm{v}}+\mathcal{L}_{\textrm{t}}\right)\rho_{s}(t),\label{eq:TME}
\end{equation}
which describes the dynamics of the reduced density matrix (RDM) $\rho_{s}$
of the three-level system. The Liouville operator $\mathcal{L}_{0}$
denotes the non-dissipative term, $\mathcal{L}_{\textrm{p}}$, $\mathcal{L}_{\textrm{v}}$
and $\mathcal{L}_{\textrm{t}}$ denote the dissipation effects associated
with the pumping, phonon coupling, and trapping, respectively.

To quantitatively investigate the energy transfer process, we define
the steady state energy fluxes by calculating the energy change of
the three-level system: 
\begin{eqnarray}
\dot{E}(\infty) & = & \mbox{Tr}_{s}[\frac{d\rho_{s}}{dt}H_{s}]\vert_{t=\infty}=\sum_{i=\mbox{p,}\mbox{v,}\mbox{t}}\mbox{Tr}_{s}\left[\mathcal{L}_{i}\left[\rho_{s}(\infty)\right]H_{s}\right]\nonumber \\
 & \equiv & \mathcal{J}_{\textrm{p}}+\mathcal{J}_{\textrm{v}}+\mathcal{J}_{\textrm{t}}.\label{eq:def_flux}
\end{eqnarray}
It can be shown that $\mbox{Tr}_{s}\left[\mathcal{L}_{0}H_{s}\right]=0$.
The three energy fluxes $\mathcal{J}_{i}$, $i=\textrm{p},\textrm{v},\textrm{t}$
are defined with respect to their corresponding dissipation operator
$\mathcal{L}_{i}$. These energy fluxes have clear physical meanings
of the energy exchange rate with the pumping field, phonon environment,
and trapping field, respectively. In this work, we are interested
in the steady state, in Eq.(\ref{eq:def_flux}) the fluxes are
calculated with $\rho_{s}(\infty)$, which is obtained by
solving $\dot{\rho}_{s}(t)=0$. Straightforwardly, we define the energy
transfer efficiency by
\begin{equation}
\eta=\left|\frac{\mathcal{J}_{\textrm{t}}(\infty)}{\mathcal{J}_{\textrm{p}}(\infty)}\right|,\label{eq:def_effi}
\end{equation}
which is the ratio between the output and the input energy fluxes. 

Without losing generality, we assume the pumping (trapping) bath is
weakly coupled with the system and can be described phenomenologically
by the local Liouville operator of the Lindblad form: 
\begin{eqnarray}
\mathcal{L}_{i}[\rho_{s}] & = & \frac{\gamma_{i}}{2}[(n_{i}+1)\left(2O_{i}^{-}\rho_{s}O_{i}^{+}-\left\{ O_{i}^{+}O_{i}^{-},\rho_{s}\right\} \right)\nonumber \\
 &  & +n_{i}\left(2O_{i}^{+}\rho_{s}O_{i}^{-}-\left\{O_{i}^{-}O_{i}^{+},\rho_{s}\right\} \right)],\label{eq:Lindblad}
\end{eqnarray}
where $i=\textrm{p},\textrm{t}$ refers to the two photon baths, $\gamma_{i}$
and $n_{i}$ are the corresponding decay rate and average photon number,
and the system operators are defined as $O_{\textrm{p}}^{+}=\left|1\right\rangle \left\langle 0\right|$, 
$O_{\textrm{t}}^{+}=\left|2\right\rangle \left\langle 0\right|$.
The system-phonon bath coupling will be treated more rigorously as
we are interested in how this coupling affects the energy transfer
over a broad range. To achieve this goal, we apply the PTRE equation,
which will be introduced in the following section.

\section{polaron-transformed Redfield equation (PTRE)}

\subsection{Secular-Markovian Redfield equation in the polaron frame}

The Redfield master equation is valid up to the second order perturbation
of the system-bath interaction. In order to go beyond this weak coupling
limit, polaron transformation is introduced to incorporate the high-order
system-bath interaction into the dynamics of the system. Here we focus
on the coupling strength between the system and phonon bath, and the
polaron transformation is only related to the two excited states.
Therefore, it is convenient to consider the dissipative dynamics of
the TLS first, then the resulting Liouville operator describing the TLS dissipative
process can be incorporated into the three-level system dynamics. We
employ the Pauli matrix $\sigma_{x}=\left|1\right\rangle \left\langle 2\right|+\left|2\right\rangle \left\langle 1\right|$
and $\sigma_{z}=\left|1\right\rangle \left\langle 1\right|-\left|2\right\rangle \left\langle 2\right|$, and define the polaron transformation
\begin{eqnarray}
\tilde{H}_{e} & = & e^{-i\sigma_{z}B/2}H_{e}e^{i\sigma_{z}B^{\dagger}/2}=\tilde{H}_{0}+\tilde{H}_{B}+\tilde{V},\label{eq:He_tilde}
\end{eqnarray}
where $H_{e}=H_{0}+H_{\textrm{v}}$ is the Hamiltonian of the TLS
with the phonon bath, the collective bath operator is $B=2i\sum_{k}\left(f_{k}b_{k}^{\dagger}-f_{k}^{\ast}b_{k}\right)/\omega_{\textrm{v}k}$,
and
\begin{eqnarray}
\tilde{H}_{0} & = & \frac{\epsilon}{2}\sigma_{z}+\frac{J}{2}\kappa\sigma_{x},\label{eq:Hs}\\
\tilde{H}_{B} & = & \sum_{k}\omega_{\textrm{v}k}b_{k}^{\dagger}b_{k}-\sum_{k}\frac{\left|f_{k}\right|^{2}}{\omega_{\textrm{v}k}},\label{eq:Hb}\\
\tilde{V} & = & \frac{J}{2}\left[\sigma_{x}\left(\cos B-\kappa\right)+\sigma_{y}\sin B\right].\label{eq:V}
\end{eqnarray}
The transformed system-bath interaction is $\tilde{V}$,
where $\epsilon=\epsilon_{1}-\epsilon_{2}$, the expectation value
of the bath operator $\kappa=\mbox{Tr}_{b}\left[\rho_{b}\cos B\right]$
is subtracted as a renormalization factor, and $\rho_{b}$ is the
thermal state of phonon bath. The spectrum function is chosen to be
super-Ohmic as $J\left(\omega\right)=4\pi\sum_{k}\left|f_{k}\right|^{2}\delta\left(\omega-\omega_{k}\right)=\alpha\pi\omega^{3}\omega_{c}^{-2}e^{-\omega/\omega_{c}}$,
where $\omega_{c}$ is the cut-off frequency and $\alpha$ is a dimensionless
parameter characterizing the system-bath coupling which is proportional
to $\lambda/\omega_{c}$ ($\lambda$ is the reorganization energy).
Therefore we can obtain 
\begin{eqnarray}
\kappa & = & \exp\left[-\int_{0}^{\infty}d\omega\frac{J\left(\omega\right)}{\pi\omega^{2}}\left(n_{\textrm{v}}\left(\omega\right)+\frac{1}{2}\right)\right]\nonumber \\
 & = & \exp\left\{ \frac{\alpha}{2}\left[1-\frac{2}{\left(\beta_{\textrm{v}}\omega_{c}\right)^{2}}\psi_{1}\left(\frac{1}{\beta_{\textrm{v}}\omega_{c}}\right)\right]\right\} ,\label{eq:kappa}
\end{eqnarray}
where $n_{\textrm{v}}(\omega)=[\exp(\beta_{\textrm{v}}\omega)-1]^{-1}$
and $\psi_{1}\left(x\right)=\sum_{n=0}^{\infty}\left(n+x\right)^{-2}$
is the trigamma function. 

Since the thermal average of $\tilde{V}$ is zero, then $\tilde{V}$
is of the order of bath fluctuations and is a reliable perturbation
parameter. Based on this consideration, the Born-Markov approximation
is applied to derive the PTRE for TLS in the Schrodinger picture as:
\begin{eqnarray}
\frac{d\tilde{\rho}_{e}}{dt} & = & -i\left[\tilde{H}_{0},\tilde{\rho}_{e}\right]-\sum_{\alpha,\beta=z,\pm}[\Gamma_{\alpha\beta}^{+}\tau_{\alpha}\tau_{\beta}\tilde{\rho}_{e}+\Gamma_{\beta\alpha}^{-}\tilde{\rho}_{e}\tau_{\beta}\tau_{\alpha}\nonumber \\
 &  & -\Gamma_{\beta\alpha}^{-}\tau_{\alpha}\tilde{\rho}_{e}\tau_{\beta}-\Gamma_{\alpha\beta}^{+}\tau_{\beta}\tilde{\rho}_{e}\tau_{\alpha}].\label{eq:PTRE}
\end{eqnarray}
Here, $\tilde{\rho}_{e}$ is the RDM of the TLS in the polaron frame,
and we use a new set of Pauli matrix $\tau_{\alpha}$ with respect
to the eigenstates of the Hamiltonian $\tilde{H}_{0}=\epsilon_{+}\left|+\right\rangle \left\langle +\right|+\epsilon_{-}\left|-\right\rangle \left\langle -\right|$:
\begin{eqnarray}
 &  & \tau_{z}=\left|+\right\rangle \left\langle +\right|-\left|-\right\rangle \left\langle -\right|,\label{eq:tau_z}\\
 &  & \tau_{+}=\left|+\right\rangle \left\langle -\right|,\ \tau_{-}=\left|-\right\rangle \left\langle +\right|.\label{eq:tau_pm}
\end{eqnarray}
The corresponding eigenvalues and eigenstates are defined by
\begin{eqnarray}
\epsilon_{\pm} & = & \pm\frac{1}{2}\sqrt{\epsilon^{2}+\left(\kappa J\right)^{2}},\label{eq:e_pm}\\
\left|+\right\rangle  & = & \cos\frac{\theta}{2}\left|1\right\rangle +\sin\frac{\theta}{2}\left|2\right\rangle ,\label{eq:+}\\
\left|-\right\rangle  & = & \sin\frac{\theta}{2}\left|1\right\rangle -\cos\frac{\theta}{2}\left|2\right\rangle ,\label{eq:-}
\end{eqnarray}
with $\tan\theta=\kappa J/\epsilon$. The transition rates $\Gamma_{\alpha\beta}^{\pm}$
are related to the half-side Fourier transformation of the bath correlation
functions
\begin{eqnarray}
\Gamma_{\alpha\beta}^{\pm}  =\frac{J^{2}}{4} \int_{0}^{\infty}dt\left\langle \xi_{\alpha}\left(\pm t\right)\xi_{\beta}\left(0\right)\right\rangle ,\label{eq:corre}
\end{eqnarray}
with
\begin{eqnarray}
\xi_{z}\left(t\right) & = & \sin\theta\left(\cos B\left(t\right)-\kappa\right),\label{eq:D8-1}\\
\xi_{\pm}\left(t\right) & = & -e^{\pm i\Delta t}\left[\cos\theta\left(\cos B\left(t\right)-\kappa\right)\mp i\sin B\left(t\right)\right].
\end{eqnarray}
The PTRE was firstly introduced by Silbey and coworkers \cite{Silbey1971,Silbey1984},
and has been widely used in solving the strong system-bath coupling
problems. Moreover, it will be shown in Sec. IV that the results
given by PTRE are consistent with those given by the Redfield equation
in the weak coupling limit and the Fermi's golden rule (or F{\"o}rster
theory) in the strong coupling limit \cite{CWang2014,Cao2015,Cao2012}.
Therefore, the PTRE smoothly connects the two limits, and provides
a useful tool to study the intermediate coupling region where there
are usually no reliable approximation methods.

\subsection{Steady state of PTRE}

For convenience, we rewrite Eq.(\ref{eq:PTRE}) in the form of the
Bloch equation
\begin{equation}
\frac{d}{dt}\left\langle \vec{\tau}\left(t\right)\right\rangle _{e}=-M\left\langle \vec{\tau}\left(t\right)\right\rangle _{e}+\vec{C}.\label{eq:B_eq-1}
\end{equation}
Here $\left\langle \vec{\tau}\left(t\right)\right\rangle _{e}^{T}=[\left\langle \tau_{z}\left(t\right)\right\rangle _{e},\left\langle \tau_{x}\left(t\right)\right\rangle_{e},\left\langle \tau_{y}\left(t\right)\right\rangle_{e}]$
with $\left\langle \cdot\right\rangle _{e}=\mbox{Tr}_{s}[\tilde{\rho}_{e}\left(t\right)\cdot]$
are the elements of the density matrix $\tilde{\rho}_{e}\left(t\right)$,
which are written in the form of the average values of the Pauli operators.
The transition matrix $M$ and the constant term $\vec{C}^{T}=\left(C_{z},C_{x},C_{y}\right)$
are
\begin{eqnarray}
M & = & \left(\begin{array}{ccc}
\gamma_{z} & \gamma_{zx} & 0\\
\gamma_{xz} & \gamma_{x} & \Delta+\gamma_{xy}\\
\gamma_{yz} & -\Delta+\gamma_{yx} & \gamma_{y}
\end{array}\right),\label{eq:M}\\
\vec{C}^{T} & = & \left(C_{z},C_{x},C_{y}\right),\label{eq:C}
\end{eqnarray}
where the eigenenergy level spacing is $\Delta=\epsilon_{+}-\epsilon_{-}$
and the expressions of the matrix elements are given in Appendix A.

\begin{figure}
\includegraphics[width=8cm]{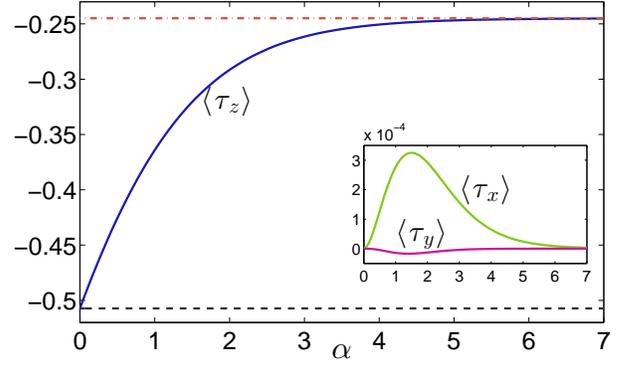}

\protect\caption{\label{fig:SS} (color online) The steady state of TLS as a function
of the system-bath coupling strength $\alpha$. The steady states
of the PTRE follow the canonical distribution in the polaron transformed
basis, which rotates with the coupling strength $\alpha$. In the
weak coupling limit, the system steady state is the canonical distribution
in the eigen basis (black dash line); while in the strong coupling
limit, the steady state is the canonical distribution in the localized
basis (red dot-dash line). The inset shows the coherent term of the
steady state, which is small in the polaron transformed basis. We
choose the parameters in units of $J$: $\epsilon_{1}/J=5$, $\epsilon_{2}/J=4.5$,
$\omega_{c}/J=5$ and $\beta_{\textrm{v}}J=1$.}
\end{figure}

The time evolution of $\left\langle \vec{\tau}\left(t\right)\right\rangle_{e}$
is straightforwardly given by
\begin{equation}
\left\langle \vec{\tau}\left(t\right)\right\rangle _{e}=e^{-Mt}\left[\left\langle \vec{\tau}\left(0\right)\right\rangle _{e}-M^{-1}\vec{C}\right]+M^{-1}\vec{C},\label{eq:TLSeqPTRE}
\end{equation}
with the steady state $\left\langle \vec{\tau}\left(\infty\right)\right\rangle _{e}=M^{-1}\vec{C}$,
and in the following we will neglect time argument $\infty$ when
referring to the steady state for convenience. The population difference
$\left\langle \tau_{z}\right\rangle_{e}$ varies with the coupling
strength as shown in Fig.\ref{fig:SS}. In the weak coupling limit,
the TLS steady state distribution is canonical with respect to its
eigen basis, i.e., 
\begin{equation}
\lim_{\alpha\rightarrow0}\left\langle \tau_{z}\right\rangle _{e}=\frac{1-\exp(\beta_{\textrm{v}}\sqrt{\epsilon^{2}+J^{2}})}{1+\exp(\beta_{\textrm{v}}\sqrt{\epsilon^{2}+J^{2}})},\label{eq:limTau_z}
\end{equation}
which is also the result $\rho_{e}^{\textrm{can}}\sim\exp(-\beta_{\textrm{v}}H_{0})$
given by equilibrium thermodynamics. When the system-bath coupling
gradually increases, the system distribution deviates from $\rho_{e}^{\textrm{can}}$
and follows the Boltzmann distribution 
\begin{equation}
\left\langle \tau_{z}\right\rangle _{e}=\frac{1-\exp(\beta_{\textrm{v}}\sqrt{\epsilon^{2}+\kappa^{2}J^{2}})}{1+\exp(\beta_{\textrm{v}}\sqrt{\epsilon^{2}+\kappa^{2}J^{2}})},\label{eq:BoltzTau_z}
\end{equation}
with respect to the eigenenergy in the polaron transformed basis $\left|+\right\rangle$
and $\left|-\right\rangle$. In the strong coupling limit, we have
\begin{equation}
\lim_{\alpha\rightarrow\infty}\left\langle \tau_{z}\right\rangle _{e}=\frac{1-\exp(\beta_{\textrm{v}}\epsilon)}{1+\exp(\beta_{\textrm{v}}\epsilon)},\label{eq:limTau_2}
\end{equation}
which is the Boltzmann distribution with respect to the site energies
$\epsilon_{1}$ and $\epsilon_{2}$ of the localized basis $\left|1\right\rangle$
and $\left|2\right\rangle$. The deviation from the canonical state
$\rho_{e}^{\textrm{can}}$ due to the strong system-bath coupling
has been studied via the cumulant expansion method in polaron transformed
thermodynamic distribution \cite{JSCso2012,Cao2012} and  from
the view point of energy shell deformation \cite{Dong2007,Dong2010,Xu2014}.

\subsection{Dissipative dynamics of the three-level system in the local basis}

Via the PTRE we can obtain a rather accurate description of the TLS
over a wide range of system-bath coupling strength. For further discussion
on the property of the entire three-level system with the other two
weakly coupled photon baths, all the observable quantities should
be manipulated in the same frame of reference. To achieve this goal,
we transform back into the frame of reference in the local basis.
The diagonal terms of the system RDM are easy to deal with, as $\sigma_{z}$
commutes with the polaron-transformation, 
\begin{eqnarray}
\left\langle \sigma_{z}\left(t\right)\right\rangle  & = & \mbox{Tr}_{s+b}\left[\rho_{tot}\left(t\right)\sigma_{z}\right]=\mbox{Tr}_{s}\left[\rho_{e}\left(t\right)\sigma_{z}\right]\nonumber \\
 & = & \mbox{Tr}_{s+b}\left[\tilde{\rho}_{tot}\left(t\right)\sigma_{z}\right]=\mbox{Tr}_{s}\left[\tilde{\rho}_{e}\left(t\right)\sigma_{z}\right],\label{eq:sgm_z}
\end{eqnarray}
where $\rho_{tot}\left(t\right)$ is the total density matrix for
both the TLS and its bath, $\rho_{e}\left(t\right)=\mbox{Tr}_{b}\left[\rho_{tot}\left(t\right)\right]$,
and $\tilde{\rho}_{tot}\left(t\right)=e^{-i\sigma_{z}B/2}\rho_{tot}\left(t\right)e^{i\sigma_{z}B^{\dagger}/2}$
is the polaron-transformed total density matrix. However, the polaron
transformation operator and $\sigma_{x}(\sigma_{y})$ do not commute
with each other, thus the off-diagonal terms cannot be obtained exactly.
We can use the approximation $\tilde{\rho}_{tot}\left(t\right)\approx\tilde{\rho}_{e}\left(t\right)\otimes\tilde{\rho}_{b}$
to obtain meaningful expressions for $\left\langle \sigma_{x}\left(t\right)\right\rangle $
and $\left\langle \sigma_{y}\left(t\right)\right\rangle$. This approximation
is essentially the Born approximation, which has already been used
in deriving the PTRE. Based on these arguments, we have
\begin{eqnarray}
\left\langle \sigma_{x}\left(t\right)\right\rangle  & = & \kappa\mbox{Tr}_{s}\left[\tilde{\rho}_{e}\left(t\right)\sigma_{x}\right],\nonumber \\
\left\langle \sigma_{y}\left(t\right)\right\rangle  & = & \kappa\mbox{Tr}_{s}\left[\tilde{\rho}_{e}\left(t\right)\sigma_{y}\right].\label{eq:sgm_y}
\end{eqnarray}

According to Eqs.(\ref{eq:+})(\ref{eq:-}), the Bloch vector $\left\langle \vec{\sigma}\left(t\right)\right\rangle =\left[\left\langle \sigma_{x}\left(t\right)\right\rangle, \left\langle \sigma_{y}\left(t\right)\right\rangle, \left\langle \sigma_{z}\left(t\right)\right\rangle \right]$
defined in the local basis of the TLS can be expressed with the quantities
calculated in the polaron frame as 
\begin{eqnarray}
\left\langle \sigma_{z}\left(t\right)\right\rangle  & = & \cos\theta\left\langle \tau_{z}\left(t\right)\right\rangle_{e}+\sin\theta\left\langle \tau_{x}\left(t\right)\right\rangle_{e},\label{eq:sigma_z}\\
\left\langle \sigma_{x}\left(t\right)\right\rangle  & = & \kappa\sin\theta\left\langle \tau_{z}\left(t\right)\right\rangle_{e}-\kappa\cos\theta\left\langle \tau_{x}\left(t\right)\right\rangle_{e},\label{eq:sigma_x}\\
\left\langle \sigma_{y}\left(t\right)\right\rangle  & = & -\kappa\left\langle \tau_{y}\left(t\right)\right\rangle_{e}.\label{eq:sigma_y}
\end{eqnarray}
Following from Eqs.(\ref{eq:B_eq-1})(\ref{eq:sigma_z})(\ref{eq:sigma_x})(\ref{eq:sigma_y}),
the equations of motion for the TLS can be written in the form of
$[\dot{\rho}_{e}(t)]_{ij}=\sum_{mn}[\mathcal{L}_{\textrm{v}}]_{(ij,mn)}[\rho_{e}(t)]_{mn}$,
then the expressions for the Liouville operator $\mathcal{L}_{\textrm{v}}$
are straightforwardly obtained.

The equations of motion of the three-level system are derived based
on Eq.(\ref{eq:TME}). The Liouville operator $\mathcal{L}_{\textrm{v}}$
with polaron effects has been obtained from the PTRE of the TLS. One
thing should be noted is that in the TLS, the conservation of population
gives $[\rho_{e}(t)]_{11}+[\rho_{e}(t)]_{22}=1$, while in the three-level
system the conservation relation becomes $\rho_{00}(t)+\rho{}_{11}(t)+\rho_{22}(t)=1$,
where $\rho_{ij}(t)=\left\langle i\right|\rho_{s}(t)\left|j\right\rangle $.
The effects of the pumping and trapping baths are described by the
Lindblad operator $\mathcal{L}_{\textrm{p}}$ and $\mathcal{L}_{\textrm{t}}$
defined in Eq.(\ref{eq:Lindblad}). Therefore, the PTRE for the three-level
system is given as 
\begin{eqnarray}
\frac{d}{dt}\left(\begin{array}{c}
\rho_{11}(t)-\rho_{22}(t)\\
\rho_{11}(t)+\rho_{22}(t)\\
\Re\left[\rho_{12}(t)\right]\\
\Im\left[\rho_{12}(t)\right]
\end{array}\right)&=&-\bar{M}\left(\begin{array}{c}
\rho_{11}(t)-\rho_{22}(t)\\
\rho_{11}(t)+\rho_{22}(t)\\
\Re\left[\rho_{12}(t)\right]\\
\Im\left[\rho_{12}(t)\right]
\end{array}\right)\nonumber\\
&+&\frac{J^{2}}{4}\left(\begin{array}{c}
\gamma_{\textrm{p}}n_{\textrm{p}}-\gamma_{\textrm{t}}n_{\textrm{t}}\\
\gamma_{\textrm{p}}n_{\textrm{p}}+\gamma_{\textrm{t}}n_{\textrm{t}}\\
0\\
0
\end{array}\right).\label{eq:BE_3LS}
\end{eqnarray}
The matrix $\bar{M}$ is shown in Appendix B. The equations for the
off-diagonal terms $\rho_{01}(t)$ and $\rho_{02}(t)$ are decoupled
from Eq.(\ref{eq:BE_3LS}) and not related with the energy flux and
transfer efficiency; thus $\rho_{01}(t)$ and $\rho_{02}(t)$ will
not be involved in the following discussion.

\section{Energy transfer flux and efficiency}

\subsection{Steady state flux}

\begin{figure}
\includegraphics[width=8cm]{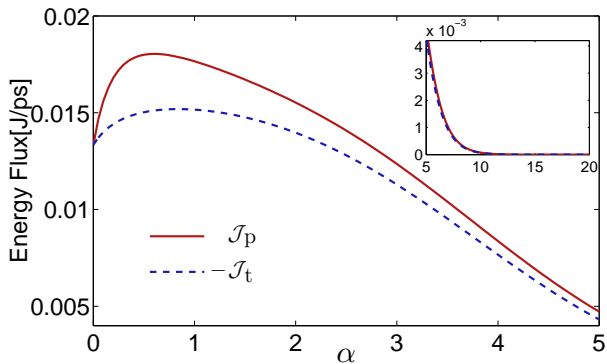}\protect\caption{\label{fig:Flux} (color online) The steady state pumping (red solid
line) and trapping (blue dashed line) energy fluxes versus $\alpha$.
The minus sign in front of the trapping flux suggests the energy
flows into the trapping bath. Both fluxes show a maximal value in
the weak coupling case and then quickly decreases to zero when $\alpha$
increases. The inset shows the strong coupling case. We use the same
parameters of the two level-system as in Fig.\ref{fig:SS}, and
the other parameters are chosen as: $\beta_{\textrm{p}}J=0.02$, $\beta_{\textrm{t}}J=1$,
and $\gamma_{\textrm{p}}/J=\gamma_{\textrm{t}}/J=0.01$. }
\end{figure}

The steady state of the three-level system can be easily obtained
from Eq.(\ref{eq:BE_3LS}), which incorporates the polaron effects
of the phonon bath. Then the steady state energy fluxes defined in
Eq.(\ref{eq:def_flux}) are straightforwardly given as 
\begin{eqnarray}
\mathcal{J}_{\textrm{p}}  =  \epsilon_{1}\gamma_{\textrm{p}}[n_{\textrm{p}}\rho_{00}-(n_{\textrm{p}}+1)\rho_{11}]-\frac{J\gamma_{\textrm{p}}}{2}(n_{\textrm{p}}+1)\Re\left[\rho_{12}\right],&\ &\label{eq:J_pu}\\
\mathcal{J}_{\textrm{t}}  =  \epsilon_{2}\gamma_{\textrm{t}}\left[n_{\textrm{t}}\rho_{00}-(n_{\textrm{t}}+1)\rho_{22}\right]-\frac{J\gamma_{\textrm{t}}}{2}(n_{\textrm{t}}+1)\Re\left[\rho_{12}\right],&\ &\label{eq:J_tr}
\end{eqnarray}
where we denote the steady state elements of RDM by $\rho_{ij}=\langle i\vert\rho_{s}(\infty)\vert j\rangle$
for brevity. Fig.\ref{fig:Flux} presents energy fluxes with respect
to $\alpha$. In the extreme case that the system bath coupling is
switched off ($\alpha=0$), there is no loss of excitation energy,
which results in $\mathcal{J}_{\textrm{p}}=-\mathcal{J}_{\textrm{t}}$,
suggesting the input energy flux from the pump completely flows into
the trap through the three-level system (note that we chose the positive direction 
as that the flux flows into the system). When the coupling
turns on, a portion of energy flux leaks into the phonon bath thus
$\mathcal{J}_{\textrm{p}}>-\mathcal{J}_{\textrm{t}}$. Both the pumping
and trapping energy fluxes reach their optimal values in the intermediate
coupling region and decrease to zero when the coupling strength is
strong.

In the context of heat engine, the trapping energy flux $\mathcal{J}_{\textrm{t}}$
in our model corresponds to the output power and $\mathcal{J}_{\textrm{p}}$
corresponds to the input power. Usually, the power of a heat engine
is small when the efficiency is high. Particularly, at the maximal
efficiency, all the processes are required to be quasi-static and
take infinite time, and thus the power will be zero. To balance the
conflict between the efficiency and power, much work has been done
to study the efficiency at maximum power \cite{Curzon2005,CVBoreck2005,CVBroeck2009}.
In the following, we will calculate the energy transfer efficiency
of our system and show its competitive relation with the trapping
flux, in analogy to the efficiency and power in the heat engine.

\subsection{Steady state efficiency}

\begin{figure}
\includegraphics[width=8cm]{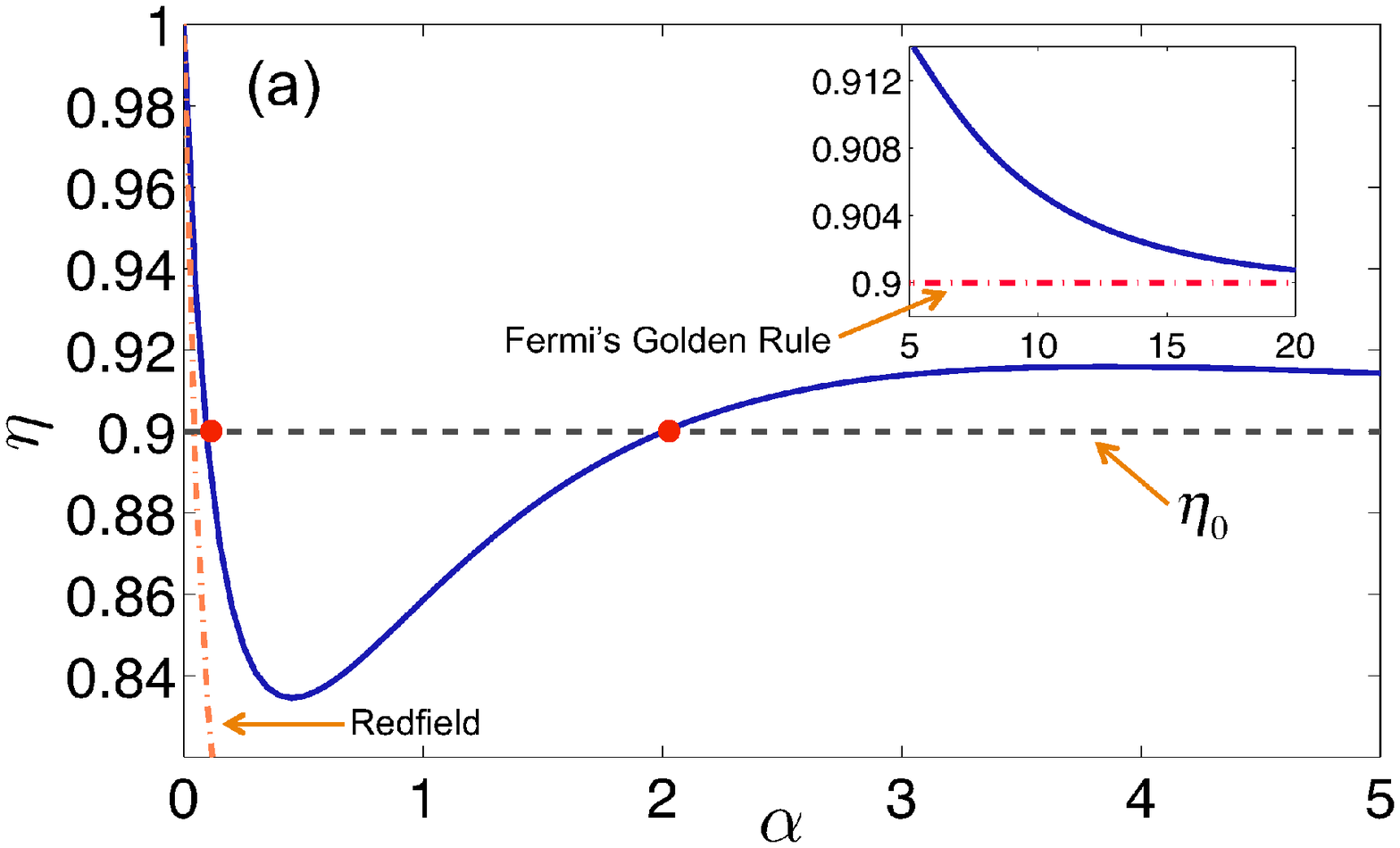}

\includegraphics[width=8cm]{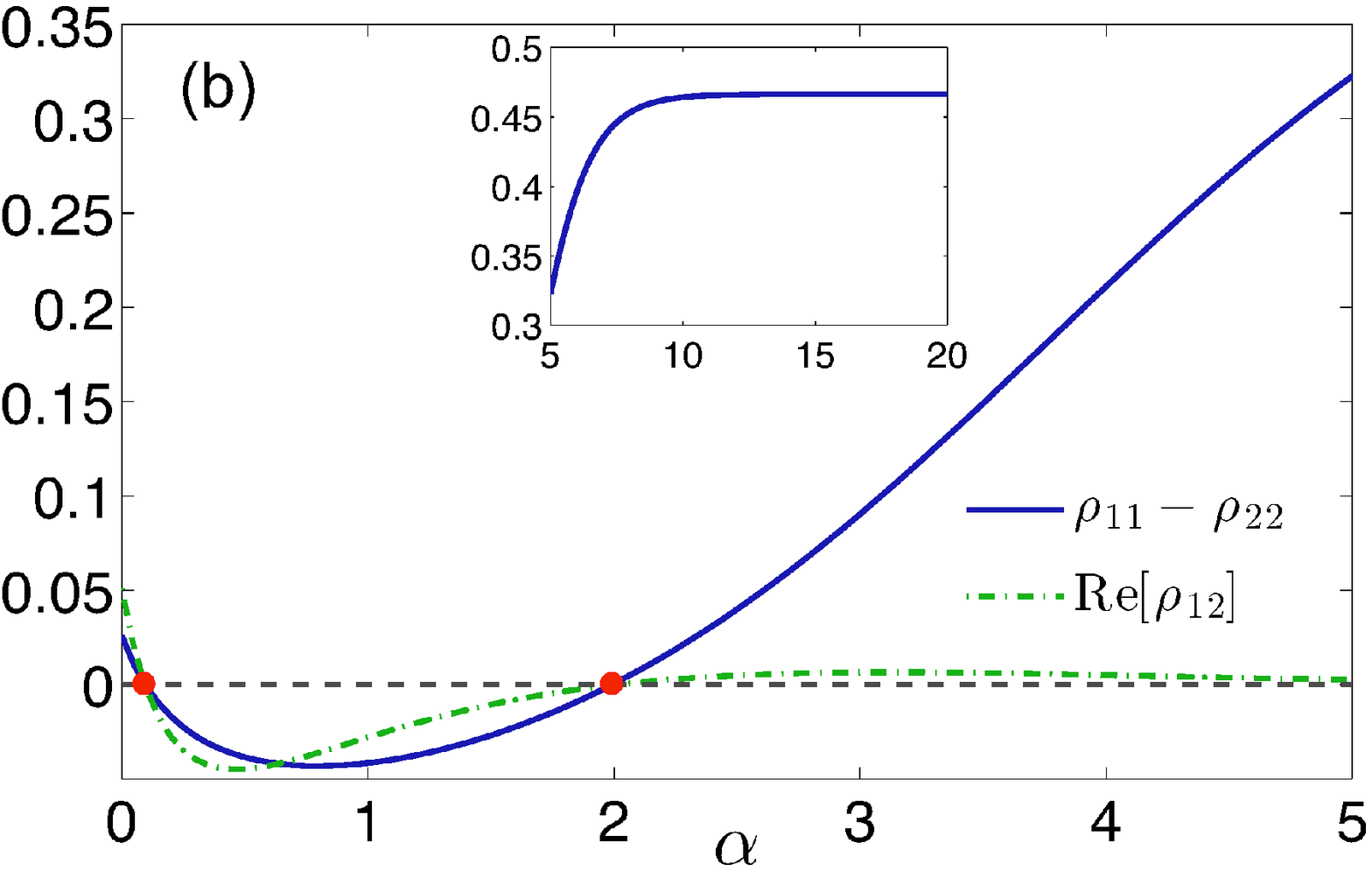}

\protect\caption{\label{fig:Effi_popu} (color online) (a) The steady states efficiency
$\eta$, (b) the excited states population $\rho_{11}-\rho_{22}$
and coherence $\Re[\rho_{12}]$ versus the system-phonon bath coupling
strength characterized by dimensionless parameter $\alpha$. The dashed
line indicates the strong coupling limit $\eta_{0}$ in (a). When
the populations are inverted, $\eta$ is less then $\eta_{0}$, the
red dots indicate the corresponding range of $\alpha$. The results
given by the Redfield equation and the Fermi's golden rule are shown
with the dashed-dot lines. The strong coupling regions are plotted
in the insets. All the parameters are chosen as the same as in Fig.\ref{fig:Flux}.}
\end{figure}

Before presenting the result of efficiency defined in Eq.(\ref{eq:def_effi}),
we begin with the analysis of the limiting cases. The first term on
the right side of Eqs.(\ref{eq:J_pu})(\ref{eq:J_tr}) depends only
on the populations of the three-level system, and the second term
represents the contribution of the off-diagonal terms (coherence in
the local basis). As we have showed in Sec III, the steady state coherence
in the local bases $\rho_{12}$ vanishes in the strong coupling limit,
then the efficiency is completely determined by the populations. According
to the steady state solution of the second equation in Eq.(\ref{eq:BE_3LS}),
we obtain the relation 
\begin{equation}
\gamma_{\textrm{p}}[n_{\textrm{p}}\rho_{00}-(n_{\textrm{p}}+1)\rho_{11}]=\gamma_{\textrm{t}}[(n_{\textrm{t}}+1)\rho_{22}-n_{\textrm{t}}\rho_{00}].\label{eq:rela}
\end{equation}
With this relation, the efficiency in the strong coupling limit reads
\begin{equation}
\eta\approx-\frac{\epsilon_{2}\gamma_{\textrm{t}}[(n_{\textrm{t}}+1)\rho_{22}-n_{\textrm{t}}\rho_{00}]}{\epsilon_{1}\gamma_{\textrm{p}}[(n_{\textrm{p}}+1)\rho_{11}-n_{\textrm{p}}\rho_{00}]}=\frac{\epsilon_{2}}{\epsilon_{1}}.\label{eq:eta_appr}
\end{equation}
This result indicates that when the coherence is negligible due to
the strong system-phonon coupling, the energy transfer efficiency
$\eta$ approaches  $\eta_{0}$, which is consistent with the key
result of Refs.\cite{Tannor2006}. We notice that Eq.(\ref{eq:rela})
shows that the net rate of pumping one excitation to $\left|1\right\rangle$
equals to the net rate of trapping one excitation from $\left|2\right\rangle$
to $\left|0\right\rangle$. In general, the efficiency is closely
related to the phonon bath induced coherence \cite{Cao2015-2} of the excited states.
If we require the system outputs positive energy, i.e., $\gamma_{\textrm{t}}(n_{\textrm{t}}+1)\rho_{22}>\gamma_{\textrm{t}}n_{\textrm{t}}\rho_{00}$,
then according to Eqs.(\ref{eq:J_pu})(\ref{eq:J_tr})(\ref{eq:rela}),
$\Re\left[\rho_{12}\right]>0$ leads to $\eta>\eta_{0}$ and vise
versa.

According to our discussion of the flux in the last subsection, when
the coupling strength $\alpha=0$, the energy transfer efficiency
$\eta=1$ because there is no loss of energy flux. When the coupling
strength gradually increases, the efficiency decreases. However, after
reaching its minimum value, the efficiency starts to rise with $\alpha$,
which is shown in Fig.\ref{fig:Effi_popu}(a). The increase of efficiency
assisted by noise was studied extensively in the context of energy transfer
in light-harvesting systems \cite{Plenio2009,Aspuru2009,Cao2009}. As we further
increase $\alpha$, the efficiency grows beyond the strong coupling
limit $\eta_{0}$ and then gradually approaches this limit from above.
The strong coupling region is plotted in the inset of Fig.\ref{fig:Effi_popu}(a).

Interestingly, we find population inversion of the two excited states
in the regimes of $\eta>\eta_{0}$. We plot the population difference
between states $\left|1\right\rangle $ and $\left|2\right\rangle$
in Fig.\ref{fig:Effi_popu}(b). In the intermediate coupling region
indicated between the two red dots, the steady state population satisfies
$\rho_{11}<\rho_{22}$ (the effective temperature associates with
these two states is positive), the corresponding efficiency $\eta$
is less then $\eta_{0}$ as shown in Fig.\ref{fig:Effi_popu}(a).
On the contrary, outside this intermediate region, i.e., when the
coupling is either very weak or very strong, the populations are inverted
$\rho_{11}>\rho_{22}$ (the effective temperature is negative); meanwhile
$\eta$ increase beyond $\eta_{0}$. In the local basis, the population
and coherence are coupled with each other due to the polaron effects:
The population inversion happens when $\Re[\rho_{12}]<0$ [Fig.\ref{fig:Effi_popu}(b)].
The fact that the population and coherence in the local basis have
similar behaviour can be explained from Eq.(\ref{eq:sigma_z}) and
Eq.(\ref{eq:sigma_x}). Here, the coherence $\left\langle \tau_{x}\left(t\right)\right\rangle _{e}$
in the polaron basis is negligibly small (see the inset of Fig.\ref{fig:SS})
to have significant effects, then the terms $\left\langle \sigma_{z}\right\rangle =\rho_{11}-\rho_{22}$
and $\left\langle \sigma_{x}\right\rangle =2\Re[\rho_{12}]$ are both
determined by $\left\langle \tau_{z}\right\rangle _{e}$.

In Fig.\ref{fig:Effi_popu}(a), we also compare the efficiency $\eta$
calculated by the PTRE method with those predicted by the Redfield
equation and the Fermi's golden rule approaches. As we mentioned before,
in the weak and strong coupling limits, the PTRE method agrees with
the Redfield equation and the Fermi's golden rule, respectively,
and it connects these two limits with a non-trivial minimum which
is related to the coherence in the local basis.

\subsection{Further discussions}

\subsubsection{kinetic models}

\begin{figure}
\includegraphics[width=8cm]{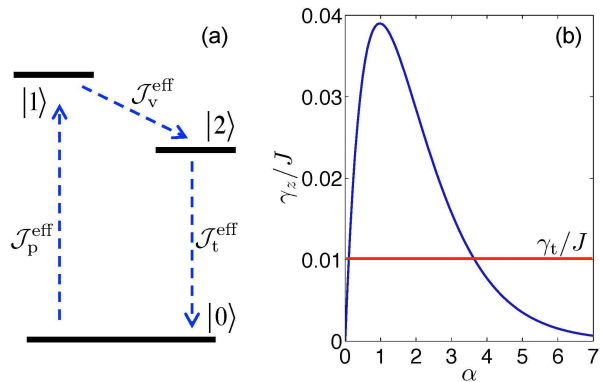}\protect\caption{\label{fig:strongcoupling}(color online) (a) The kinetic model of a single excitation
transfer cycle. In the strong coupling regime, the energy
transfer processes between different local states can be described
by the effective fluxes. (b) The transfer rate $\gamma_{z}$ versus
system-bath coupling strength $\alpha$. The red line indicates the trapping rate $\gamma_\textrm{t}$. The parameters here are the
same with Fig.\ref{fig:Flux}.}

\end{figure}

In the strong coupling regime, we can map this energy transfer process
into a simple excitation kinetic model as shown in Fig.\ref{fig:strongcoupling}(a).
Each step of energy transfer is described by an effective flux ($\mathcal{J}_{\textrm{p}}^{\textrm{eff}}$,
$\mathcal{J}_{\textrm{v}}^{\textrm{eff}}$ and $\mathcal{J}_{\textrm{t}}^{\textrm{eff}}$).
The effective transfer flux $\mathcal{J}_{\textrm{v}}^{\textrm{eff}}$
between the two excited states is approximately proportional to $\gamma_{z}$
[Fig.\ref{fig:strongcoupling}(b)], which characterizes the relaxation rate
 of the two excited states. When $\mathcal{J}_{\textrm{v}}^{\textrm{eff}}$
(or $\gamma_{z}$) is smaller than the trapping flux $\mathcal{J}_{\textrm{t}}^{\textrm{eff}}$ (or $\gamma_\textrm{t}$), 
the excitation in excited states will be quickly captured by the trapping field without
enough time to first get equilibrated with the phonon bath. 
Consequently, the populations of the two excited states are inverted and the real part of the coherence
becomes negative. This
phenomenological mechanism explains why the efficiency $\eta$ is higher than
$\eta_{0}$ in the strong coupling limit.

\begin{figure}
\includegraphics[width=8cm]{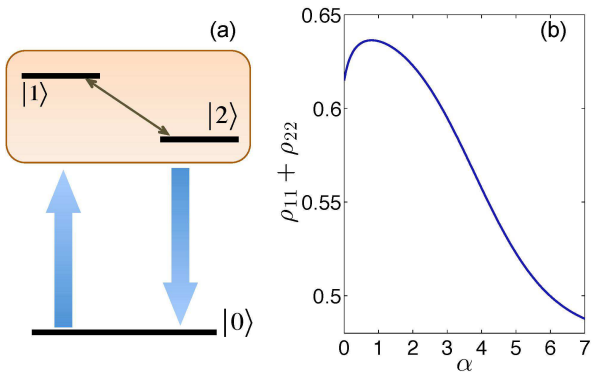}\protect\caption{\label{fig:effective} (a) The two excited states form a black box
for the input and output excitation due to the internal coupling.
(b) The average residence time $\left\langle t\right\rangle$ is
proportional to the total population of the excited states. The parameters
here are also the same as those in Fig.\ref{fig:Flux}.}
\end{figure}

When the system-bath coupling strength becomes weaker, the local basis
frame is no longer a good option for the kinetic
picture. The two excited states couple with each other and can be
together considered as an excited state manifold, as shown in Fig.\ref{fig:effective}(a).
The single excitation carrying
certain amount of energy passes through the excited states $\text{\ensuremath{\left|1\right\rangle}}$
and $\left|2\right\rangle$, and its average residence time $\left\langle t\right\rangle$
in the excited states is negatively correlated with the transfer efficiency
(in analogy to the light-harvesting efficiency in Ref.\cite{Cao2009,JSCao2013}):
i.e., the longer the excitation stays in the excited states, the more energy 
will be lost to the phonon bath, and the lower
energy transfer efficiency will be. During a cycle that the single
excitation starts from $\left|0\right\rangle$ and finally returns
to $\left|0\right\rangle$, the average residence time $\left\langle t\right\rangle$
 is proportional to the excited states population
$\rho_{11}+\rho_{22}$ at the steady states, as shown in Fig.\ref{fig:effective}(b).
Though not quantitively exact, this kinetic model qualitatively
explains the local minimal of the efficiency $\eta$ via the average
residence time $\left\langle t\right\rangle \sim\rho_{11}+\rho_{22}$.

\subsubsection{temperature dependence}

\begin{figure}
\includegraphics[width=9cm]{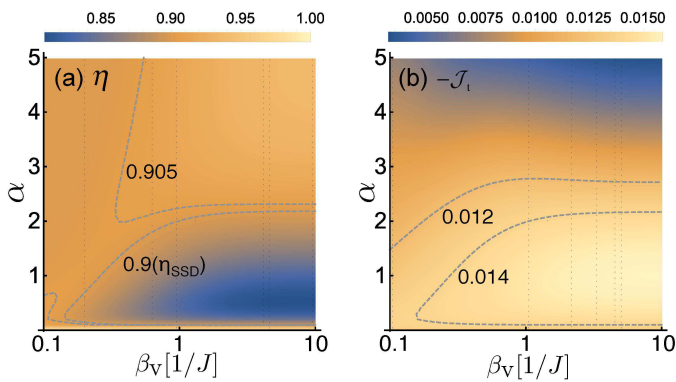}
\protect\caption{\label{fig:2D}(color online) The dependence of (a) the energy transfer
efficiency and (b) the trapping energy flux on the coupling strength
and temperature of the phonon bath. The temperatures and dissipation
coefficients of the pumping and trapping bath are the same as in the
Fig.\ref{fig:Effi_popu}. }
\end{figure}

Besides the system-phonon bath coupling strength, the temperature
of the phonon bath also affects the energy transfer process, as shown
in the two-dimensional contours of energy transfer efficiency {[}Fig.\ref{fig:2D}(a){]}
and trapping energy flux [Fig.\ref{fig:2D}(b)]. The efficiency
behaves the same at the high phonon bath temperature as in the strong coupling.
In the high temperature limit, even when the coupling strength is
weak, the efficiency is still close to $\eta_{0}$. As seen from Eq.(\ref{eq:kappa}),
in either limit $\alpha\rightarrow\infty$ or $\beta_{\textrm{v}}\rightarrow0$,
the renormalization factor $\kappa \rightarrow 0$; therefore, except
for the weak coupling and low temperature case, the efficiency $\eta$
does not change obviously. 

The trapping energy flux has a different
temperature dependences for weak and strong system-bath couplings.
The flux $-\mathcal{J}_{\textrm{t}}$ grows (goes down) with descending
$\beta_{\textrm{v}}$ when $\alpha$ is small (large). Moreover,
$-\mathcal{J}_{\textrm{t}}$ does not sensitively depend on $\beta_{\textrm{v}}$ with small $\alpha$
in contrast with the efficiency. When the coupling $\alpha$ is around
1, the flux $-\mathcal{J}_{\textrm{t}}$ changes no more than 20\%
in amplitude comparing with its maximum. 
The optimization of the efficiency and the trapping flux can be achieved in two different regimes:
1) The coupling strength is weak and the temperature of the phonon bath is high. 2) The
coupling strength is medium ($\alpha\sim2.5$) and the bath temperature is low ($\beta_{\textrm{v}}$>1).
The first regime corresponds to the high temperature classical limit, and the second regime
corresponds to low-temperature quantum regime, where bath-induced coherence enhances 
the energy transfer process.

\section{Conclusion}

In this paper we use the polaron transformed Redfield equation (PTRE)
to analyze the effects of the phonon bath on the energy transfer process in
a generic three-level model. As a quantitative method, the PTRE can reliably
describe the dependence of the steady state coherence on the system-bath
coupling strength ranging from the weak to strong coupling regime.
Our analysis shows that the steady state coherence between the two
excited states is crucial to the energy transfer efficiency. When
the effective temperature of the excited states is negative (populations are inverted), 
the coherence carries a positive real part 
and enhances the efficiency beyond
the strong coupling limit $\eta_{0}$. On the contrary, if the effective
temperature is positive (populations are not inverted), 
the coherence carries a negative real part and
is detrimental to the efficiency.
The energy flux and efficiency compete with each other and cannot
reach maximum simultaneously; however, the study of their behaviours
with respect to the coupling strength and temperature provides 
the key information about how to make an optimal compromise between
the two quantities. We will consider how to use quantum control to
optimize the energy transfer process in the future study.
\begin{acknowledgments}
DX and YZ were supported by the National Research Foundation, Republic
of Singapore, through the Competitive Research Program (CRP) under
Project No. NRF-CRP5-2009-04. JC acknowledges the National Science
Foundation (NSF) of the US (grant no. CHE-1112825). CW has been supported
by the Singapore-MIT Alliance for Research and Technology (SMART).
\end{acknowledgments}

\appendix

\section*{Appendix A $M$ and $\vec{C}$ in Eq.(\ref{eq:M}) and Eq.(\ref{eq:C})}

The quantities defined in Eq.(\ref{eq:M}) and Eq.(\ref{eq:C}) are
determined by the superposition of the correlation functions Eq.(\ref{eq:corre})
following Eqs.(\ref{eq:PTRE})(\ref{eq:B_eq-1}). The straightforwardly
calculation gives: 

\begin{eqnarray*}
\gamma_{z} & = & \frac{1}{2}\kappa^{2}J^{2}\int_{0}^{\infty}dt\cos\left(\Delta t\right)\left[f\left(t\right)\cos^{2}\theta+g\left(t\right)\right],\\
\gamma_{x} & = & \frac{1}{2}\kappa^{2}J^{2}\int_{0}^{\infty}dt\left[f\left(t\right)\sin^{2}\theta+\cos\left(\Delta t\right)g\left(t\right)\right],\\
\gamma_{y} & = & \frac{1}{2}\kappa^{2}J^{2}\int_{0}^{\infty}dtf\left(t\right)\left[\cos^{2}\theta\cos\left(\Delta t\right)+\sin^{2}\theta\right],\\
\gamma_{zx} & = & \frac{1}{4}\kappa^{2}J^{2}\sin2\theta\int_{0}^{\infty}dtf\left(t\right),\\
\gamma_{xz} & = & \frac{1}{4}\kappa^{2}J^{2}\sin2\theta\int_{0}^{\infty}dtf\left(t\right)\cos\left(\Delta t\right),\\
\gamma_{xy} & = & \frac{1}{2}\kappa^{2}J^{2}\int_{0}^{\infty}dtg\left(t\right)\sin\left(\Delta t\right),\\
\gamma_{yx} & = & -\frac{1}{2}\kappa^{2}J^{2}\cos^{2}\theta\int_{0}^{\infty}dtf\left(t\right)\sin\left(\Delta t\right),\\
\gamma_{yz} & = & \frac{1}{4}\kappa^{2}J^{2}\sin2\theta\int_{0}^{\infty}dtf\left(t\right)\sin\left(\Delta t\right),\\
C_{z} & = & -\frac{i}{2}\kappa^{2}J^{2}\int_{-\infty}^{\infty}dt\sin\left(\Delta t\right)\\
 &  & \times\left[\cos^{2}\theta\cosh\left[Q\left(t\right)\right]+\sinh\left[Q\left(t\right)\right]\right],\\
C_{x} & = & -\frac{i}{4}\kappa^{2}J^{2}\sin2\theta\int_{-\infty}^{\infty}dt\sin\left(\Delta t\right)\cosh\left[Q\left(t\right)\right],\\
C_{y} & = & -\frac{i}{4}\kappa^{2}J^{2}\sin2\theta\int_{0}^{\infty}dt\left[1-\cos\left(\Delta t\right)\right]\\
 &  & \times\left[\cosh\left[Q\left(t\right)\right]-\cosh\left[Q\left(-t\right)\right]\right],
\end{eqnarray*}
where
\begin{eqnarray*}
f\left(t\right) & = & \cosh\left[Q\left(t\right)\right]+\cosh\left[Q\left(-t\right)\right]-2,\\
g\left(t\right) & = & \sinh\left[Q\left(t\right)\right]+\sinh\left[Q\left(-t\right)\right].
\end{eqnarray*}
Using the super-Ohmic spectrum $J\left(\omega\right)=\alpha\pi\omega^{3}\omega_{c}^{-2}e^{-\omega/\omega_{c}}$,
the function $Q\left(t\right)$ reads 
\begin{eqnarray*}
Q\left(t\right) & = & \int_{0}^{\infty}d\omega\frac{J\left(\omega\right)}{\pi\omega^{2}}\left[\left(2n_{\textrm{v}}\left(\omega\right)+1\right)\cos\left(\omega t\right)-i\sin\omega t\right]\\
 & = & \alpha\left\{ \frac{-1+\omega_{c}^{2}t^{2}-i2\omega_{c}t}{\left(1+\omega_{c}^{2}t^{2}\right)^{2}}+\frac{2\Re\left[\psi_{1}\left(\frac{1}{\beta_{\textrm{v}}\omega_{c}}+\frac{it}{\beta_{\textrm{v}}}\right)\right]}{\left(\beta_{\textrm{v}}\omega_{c}\right)^{2}}\right\} .
\end{eqnarray*}

\section*{Appendix B $\bar{M}$ in Eq.(\ref{eq:BE_3LS})}

The Liouville operator $\mathcal{L}_{\textrm{v}}$ for the three-level
system is obtained from Eq.(\ref{eq:B_eq-1}) for the TLS with the
expressions in Appendix A. Here the relation $\rho_{00}+\rho_{11}+\rho_{22}=1$
should be used to substitute $[\rho_{e}]_{11}+[\rho_{e}]_{22}=1$
for the TLS. Taking the contributions of the Lindblad terms $\mathcal{L}_{\textrm{p}}$
and $\mathcal{L}_{\textrm{t}}$ defined in Eq.(\ref{eq:Lindblad})
into consideration, the elements of the matrix $\bar{M}$ in Eq.(\ref{eq:BE_3LS})
are 
\begin{eqnarray*}
\bar{M}_{11} & = & \gamma_{z}\cos^{2}\theta+\gamma_{x}\sin^{2}\theta+\frac{1}{2}(\gamma_{xz}+\gamma_{zx})\sin2\theta\\
 &  & +\frac{1}{2}[\gamma_{\textrm{p}}(n_{\textrm{p}}+1)+\gamma_{\textrm{\mbox{t}}}(n_{\textrm{t}}+1)],\\
\bar{M}_{12} & = & -C_{z}\cos\theta-C_{x}\sin\theta+\frac{1}{2}[\gamma_{\textrm{p}}(3n_{\textrm{p}}+1)-\gamma_{\textrm{\mbox{t}}}(3n_{\textrm{t}}+1)],\\
\bar{M}_{13} & = & \kappa^{-1}[\gamma_{xz}\sin^{2}\theta-\gamma_{zx}\cos^{2}\theta+\frac{1}{2}(\gamma_{z}-\gamma_{x})\sin2\theta],\\
\bar{M}_{14} & = & -\kappa^{-1}(\Delta+\gamma_{xy})\sin\theta,\\
\bar{M}_{21} & = & \frac{1}{2}[\gamma_{\textrm{p}}(n_{\textrm{p}}+1)-\gamma_{\textrm{\mbox{t}}}(n_{\textrm{t}}+1)],\\
\bar{M}_{22} & = & \frac{1}{2}[\gamma_{\textrm{p}}(3n_{\textrm{p}}+1)+\gamma_{\textrm{\mbox{t}}}(3n_{\textrm{t}}+1)],\\
\bar{M}_{23} & = & \bar{M}_{24}=0,\\
\bar{M}_{31} & = & \kappa[\gamma_{zx}\sin^{2}\theta-\gamma_{xz}\cos^{2}\theta+\frac{1}{2}(\gamma_{z}-\gamma_{x})\sin2\theta],\\
\bar{M}_{32} & = & \kappa(C_{x}\cos\theta-C_{z}\sin\theta),\\
\bar{M}_{33} & = & \gamma_{x}\cos^{2}\theta+\gamma_{z}\sin^{2}\theta-\frac{1}{2}(\gamma_{xz}+\gamma_{zx})\sin2\theta\\
 &  & +\frac{1}{2}[\gamma_{\textrm{p}}(n_{\textrm{p}}+1)+\gamma_{\textrm{\mbox{t}}}(n_{\textrm{t}}+1)],\\
\bar{M}_{34} & = & (\Delta+\gamma_{xy})\cos\theta,\\
\bar{M}_{41} & = & \kappa[(\Delta-\gamma_{yx})\sin\theta-\gamma_{yz}\cos\theta],\\
\bar{M}_{42} & = & \kappa C_{y},\\
\bar{M}_{43} & = & -(\Delta-\gamma_{yx})\cos\theta-\gamma_{yz}\sin\theta,\\
\bar{M}_{44} & = & \gamma_{y}+\frac{1}{2}[\gamma_{\textrm{p}}(n_{\textrm{p}}+1)+\gamma_{\textrm{\mbox{t}}}(n_{\textrm{t}}+1)].
\end{eqnarray*}

\end{document}